\documentclass[a4paper, oneside, twocolumn, notitlepage, 10pt]{extarticle_ecoc}
\usepackage{ecoc}
\usepackage{subcaption}

\begin{document}
\selectlanguage{english}    
\newcommand{\saip}[1]{\textit{\textcolor{blue}{[sai]: #1}}}
\newcommand{\cmas}[1]{\textit{\textcolor{magenta}{[cmas]: #1}}}

\title{Planning Optical Networks for Unexpected Traffic Growth}%


\author{
    Sai Kireet Patri\textsuperscript{(1,2)}, Achim Autenrieth\textsuperscript{(1)},
    Jörg-Peter Elbers\textsuperscript{(1)}, Carmen Mas Machuca\textsuperscript{(2)}
}

\maketitle                  


\begin{strip}
 \begin{author_descr}

   \textsuperscript{(1)} ADVA, Fraunhoferstr.~9A, 82152 Martinsried, Germany
   \textcolor{blue}{\uline{SPatri@adva.com}}

   \textsuperscript{(2)} Chair of Communication Networks, TU Munich, Arcisstr.~21, 80331 Munich, Germany

 \end{author_descr}
\end{strip}

\setstretch{1.1}


\begin{strip}
  \begin{ecoc_abstract}
    A lightpath configuration algorithm considering a multi-period traffic model is presented and evaluated to a realistic German core network study.~Strategically exploiting BVT excess capacity, we show that offered traffic can be carried up to an additional five years in all traffic growth scenarios.
  \end{ecoc_abstract}
\end{strip}
\section{Introduction}
\vspace{-0.2cm}
With 5G deployment becoming a reality across countries, core network operators are realising that deployed optical networks need to be upgraded to quickly capitalize gains.~Due to the recent pandemic, many network operators (e.g., BT and Telefónica), have reported a 30-50\% increase in broadband traffic in Q1 2020, compared to Q4 2019\cite{oecd}.~Given the worldwide impact on logistics, operators must invest in software configurable optical terminals\supercite{adva} and Bandwidth Variable Transponders~(BVTs), which offer opportunities to increase network throughput in order to carry the growth in offered traffic. We define offered traffic as the total throughput of the network if all demands are met by adding sufficient BVTs in the network.

Traditionally,~lightpath configuration in network planning consists of algorithms exploiting Integer Linear Programming~(ILP) as well as heuristics.~A link state heuristic for optical networks\supercite{Wang2019} is of particular interest; however, no insights on its usability in a multi-period scenario are provided.~The SNAP Algorithm \supercite{Virgillito2019} offers progressive traffic loading, randomly allocating traffic between nodes without considering realistic growth.~Both works also assume homogeneous spans across their physical topology, limitations of which have been discussed in our previous work\supercite{Patri20}.

In this work, we first present a traffic generator and discuss two traffic growth models \textit{viz.}, Expected and Unexpected.~Using our developed solution~(\textit{Scheme~1}) on a German core topology, we show that the offered traffic can be met by strategically leveraging the configured BVTs' capacity, hence saving bandwidth for future growth.~\textit{Scheme~1}~is evaluated against a simple baseline~(\textit{Scheme~2}),~which optimally adds new BVTs, instead of upgrading the deployed BVTs to a higher datarate when possible.~Finally, results in terms of aggregate network throughput~(carried traffic)~and provisioned BVTs are discussed.

\section{Traffic Generator and Growth Model}
\vspace{-0.1cm}
As network operators do not share their offered traffic data openly due to confidentiality, network planners have relied on traditional empirical traffic models like the gravity model\supercite{Dwivedi2000}, which assumes traffic to be proportional to the human population of the source and destination.~However, in present times, most of the traffic between two locations are exchanged between either Data Centers~(DCs) or Internet Exchange Points~(IXPs) and there exists a higher correlation of offered traffic to node degree as compared to human population. For e.g., Frankfurt and Duesseldorf are less populated cities as compared to Hamburg and Munich, but exchange traffic at higher rates\supercite{decix}. Based on these assumptions, we define initial traffic~(time~$t=0$), between source~$i$ and destination~$j$ as:
\begin{equation}\label{eq:initTrafModel}
\tau(i,j,0)[Gbps] =  
\begin{cases}
2 \cdot \binom N2 \cdot \Delta_{i} \cdot  \Delta_{j} & \text{if~$N > 2 \cdot \bar{N}$}\\
N \cdot \Delta_{i} \cdot  \Delta_{j} & \text{$otherwise$}\\
\end{cases}\;
\end{equation}
where~$\bar{N}$~is the average node degree of the topology~$ N=\sum_{ \kappa  \in \{i,j\}} N_{\kappa}$ is the combined node degree, and~$\Delta_{\kappa}$ is the absolute difference between the number of co-located DCs and IXPs at a given node~$\kappa$\supercite{dcmap}.

For~$t>0$, the offered traffic between nodes is given as~$\tau(i,j,t)=\gamma_{t} \cdot \tau(i,j,0)$, where~$\gamma_{t}$~is the expected growth in offered traffic\cite{Turkcu18}. For unexpected traffic growth, we modify $\gamma_{t}$ with an increase of 30-90\% between 2023 and 2026, as compared to its original value.~As seen in Fig.~\ref{fig:traffgrowth}, this brings about an overall realistic increase of ~40\% in the aggregate offered traffic growth for the given network topology.

\section{Algorithm and Multi-period Planning Study}

\begin{figure}[htbp!]
\centering
 \includegraphics[width=0.45\textwidth]{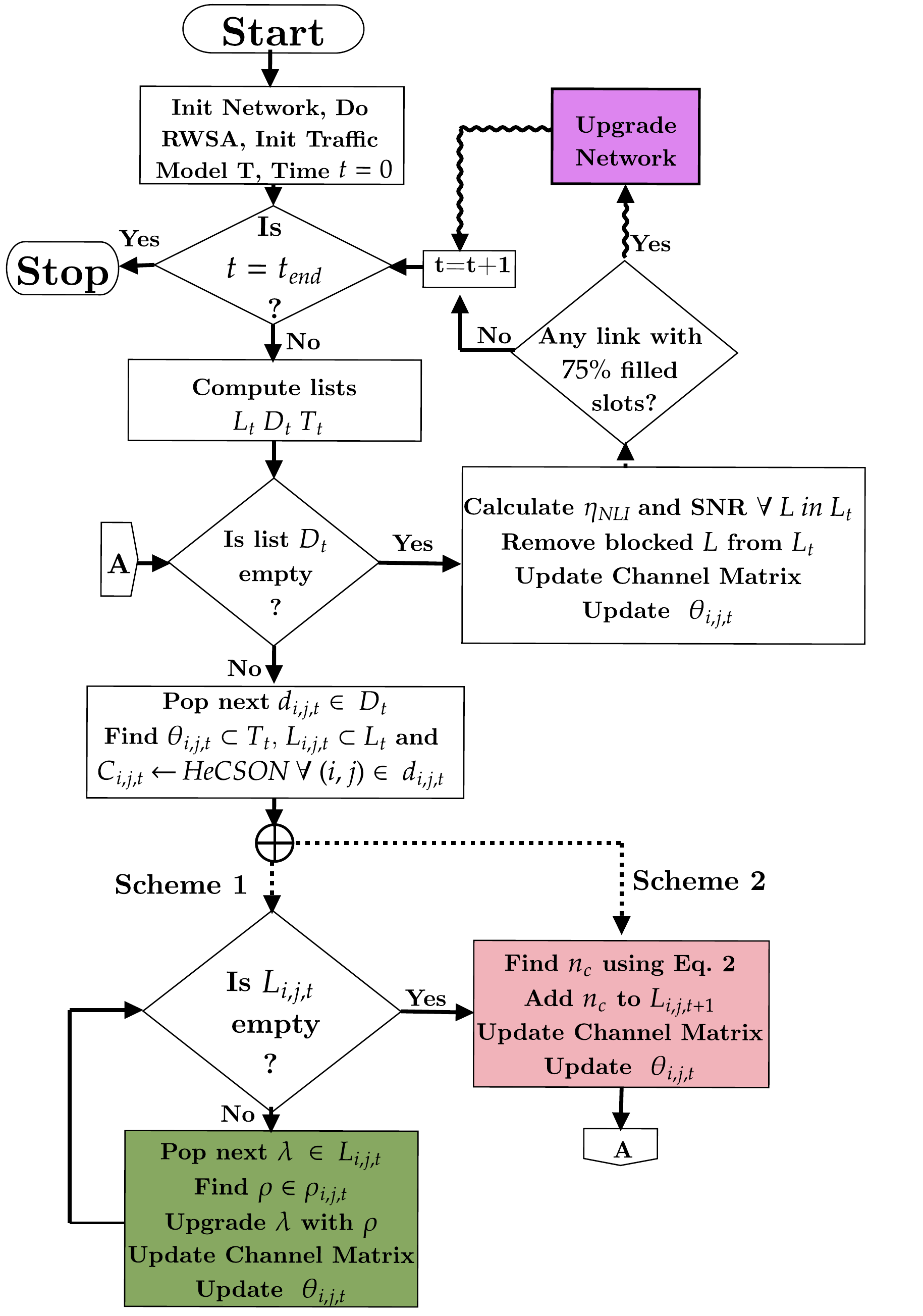}
\caption{Proposed solution flow in a multi-period scenario with two schemes and optional physical network upgrade}
\label{fig:flowchart}
\end{figure}

The multi-period planning algorithm is shown in Fig.~\ref{fig:flowchart}.~It takes as input the topology, along with information on the number of DCs and IXPs at each source destination pair~($i,j$).~We then find the initial offered traffic using Eq.~\ref{eq:initTrafModel} and also do a Routing Wavelength and Spectrum Assignment~(RWSA), using a weighted probabilistic routing based on Yen's k-Shortest Path Algorithm and the number of continuous empty frequency slots in each of the paths.~This heuristic for routing adds a randomness in choosing the candidate path list for each~$(i,j)$.~We use first-fit channel allocation strategy for spectrum assignment.~
\begin{figure*}[htbp!]
\begin{subfigure}[htbp!]{0.33\textwidth}
  \centering
  \includegraphics[width=\textwidth,height=11em]{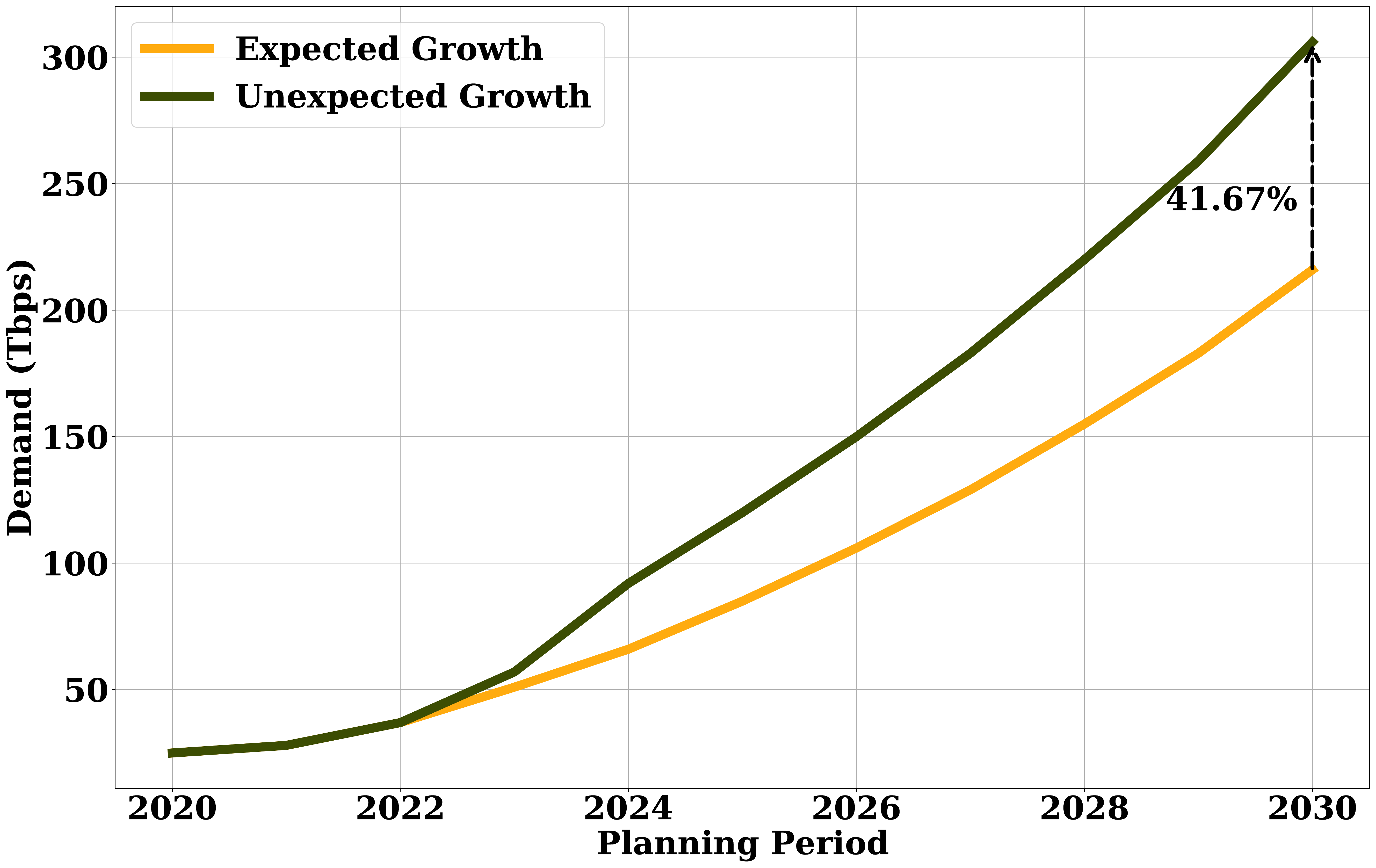}
  \caption{Aggregate Offered Traffic Growth models}
  \label{fig:traffgrowth}
\end{subfigure}
\begin{subfigure}[htbp!]{0.33\textwidth}
  \centering
  \includegraphics[width=\textwidth,height=11em]{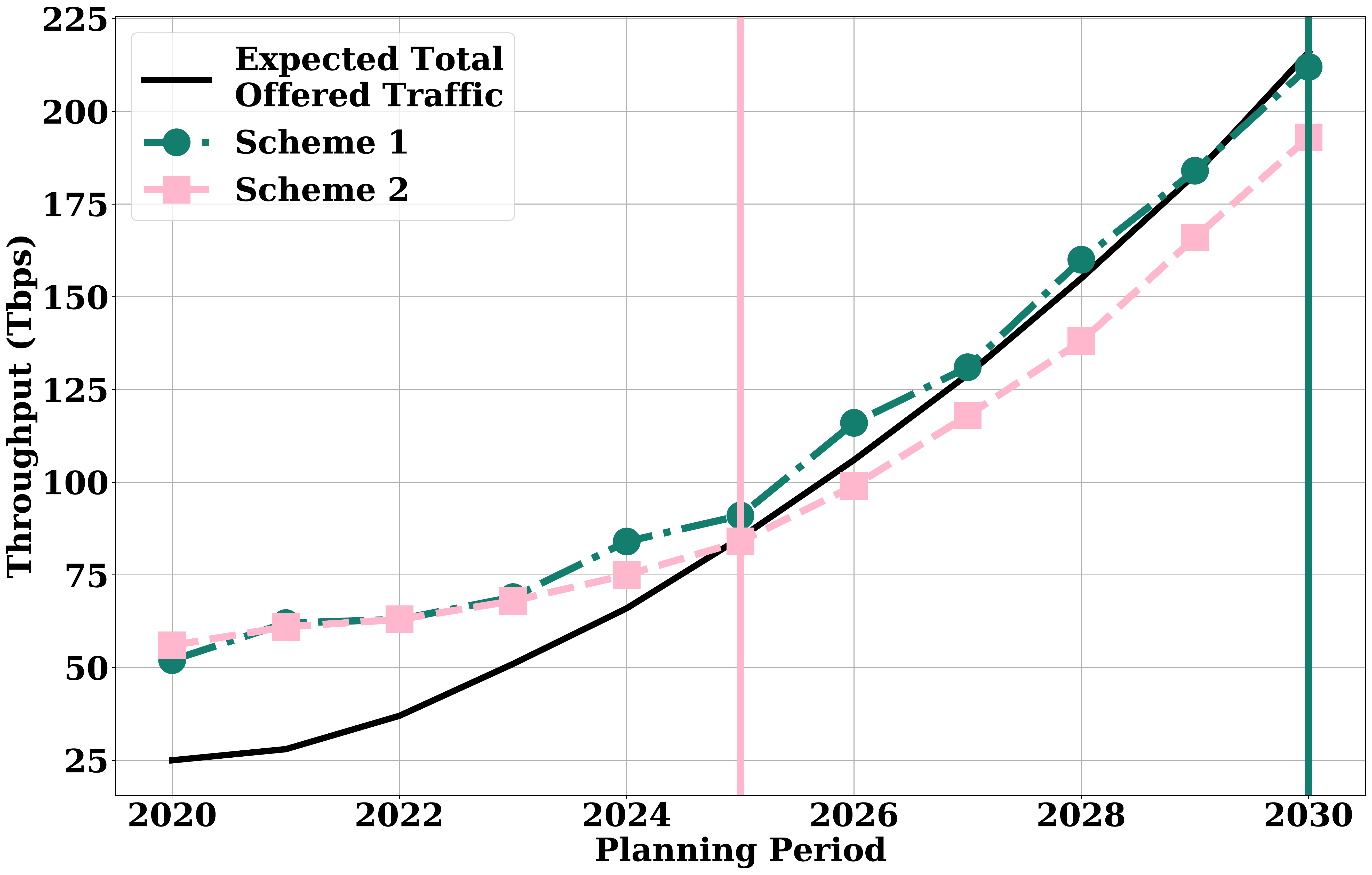}
 \caption{Expected Traffic Throughput}
  \label{fig:expgrowth}
\end{subfigure}%
\begin{subfigure}[htbp!]{0.33\textwidth}
  \centering
  \includegraphics[width=\textwidth,height=11em]{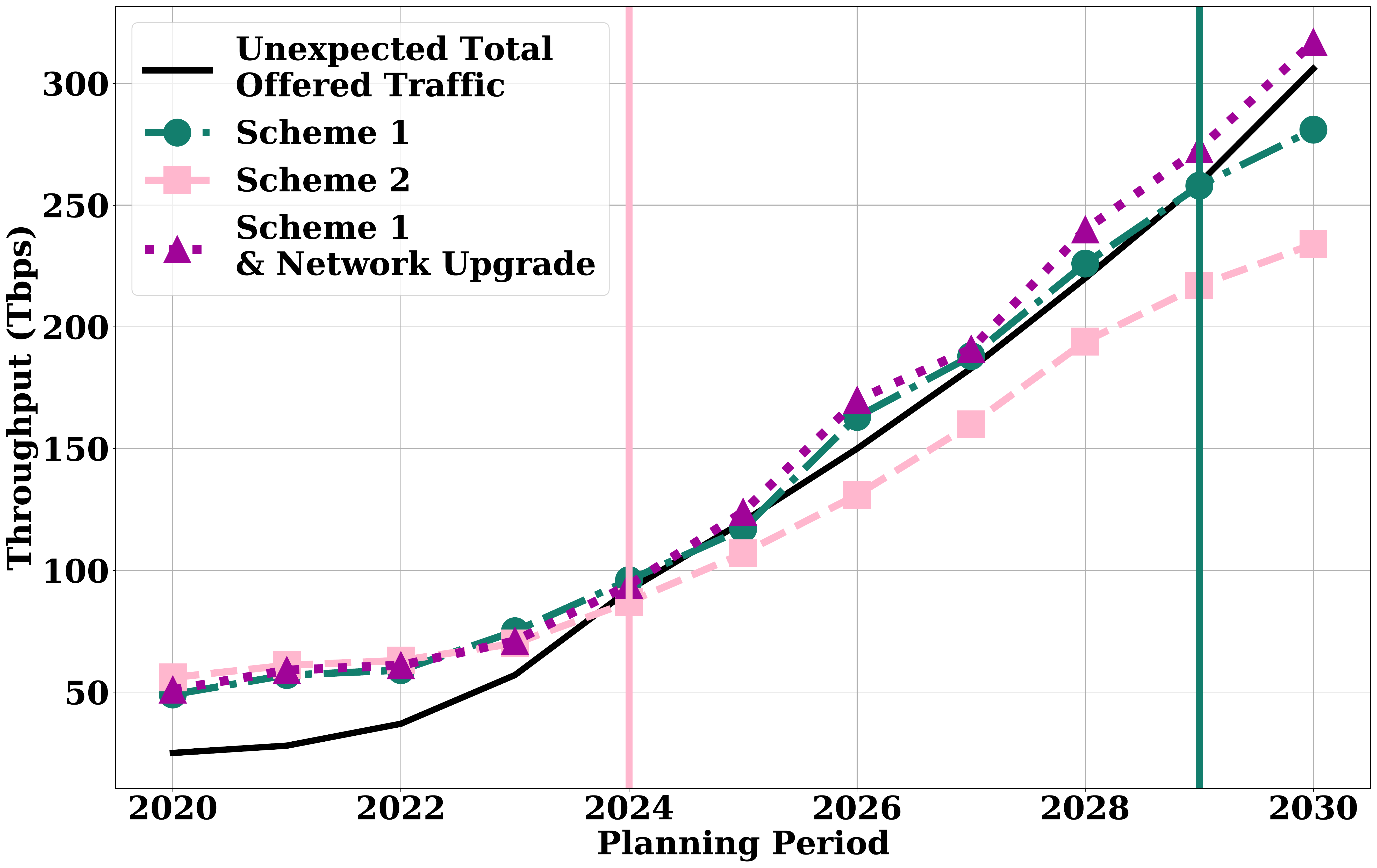}
  \caption{Unexpected Traffic Throughput}
  \label{fig:unexpgrowth}
\end{subfigure}
\vspace{1em}
\caption{Yearly aggregate throughput of Germany17 topology from 2020-2030}
\label{fig:throughput}
\end{figure*}

The objective of the algorithm is to minimize the number of lightpaths~(LPs) added to the network for planning period~$t$, while trying to meet the offered aggregate traffic. It is pertinent to note that one LP is associated to one BVT. ~To begin, we take as input the candidate path list of all demands~($D_t$), all currently provisioned LPs in the network~($L_t$) and the traffic matrix~($T_t$)~calculated using Eq.~\ref{eq:initTrafModel} and Fig.~\ref{fig:traffgrowth}.~For each candidate path $d_{i,j,t}\in D_t$, we compute a list of LPs~$L_{i,j,t}\subset L_t$ already present between nodes~$(i,j)$ and additional offered traffic~$\theta_{i,j,t}$, defined as~$\tau(i,j,t)-\tau(i,j,0)$.~Using a datarate granularity of 50 Gbps between 100-600 Gbps in steps of 50 Gbps with QAM values of QPSK, 8, 16, 32 and 64 QAM, we generate more than 60 different channel configurations~(combination of~$\{Datarate,QAM\}$)~for each BVT.~The list of valid channel configurations~$C_{i,j,t}$ is then filtered using HeCSON\cite{Patri20}.~For every provisioned LP~($\lambda \in L_{i,j,t}$), a set of channel configurations~$\rho_{i,j,t}\subset C_{i,j,t}$~is found, having their channel bandwidth~$BW_\rho$ less than or equal to the provisioned LP's channel bandwidth~$BW_\lambda$.~This ensures upgrade of provisioned LPs at the same central channel frequency, without additional bandwidth usage.~We then iterate over~$\rho_{i,j,t}$ sorted according to highest datarate~(DR) to find the first~$\rho$, which has a higher DR than~$\lambda$.~If found,~$\lambda$ is updated with~$\rho$ and the spare capacity generated can be used to satisfy some of the additional traffic, thereby reducing~$\theta_{i,j,t}$.
\begin{equation*}
\begin{array}{ll@{}ll}
minimize~~ \sum_{c~\in~C_{i,j,t}} n_{c} &\\
subject~to:~~\theta_{i,j,t}\leq \sum n_{c}\cdot DR_{c}<\theta_{i,j,t}+\delta &\\
~~~~~~~~~~0< \sum n_{c}\cdot \eta_{NLI_c}\leq\sum_{\forall \lambda \in L_{i,j,t}} \eta_{NLI_\lambda}\\
\end{array}
\label{eq:ilp}\tag{2}    
\end{equation*}
After the provisioned LPs are upgraded, we find the number of additional LPs~$n_c$ of each valid channel configuration~$c$, which are needed to carry the remaining $\theta_{i,j,t}$ using the ILP shown in Eq.~\ref{eq:ilp}.~This ILP minimizes integer~$n_c$, such that the configured total datarate can only be over-provisioned by~$\delta$ Gbps.~For our network study, we fix~$\delta$ as 100 Gbps.~We also use a power independent NLI co-efficient constraint,~$\eta_{NLI}$, following definitions and calculations of ACF-EGN model \supercite{Poggiolini19}.~This NLI 
constraint restricts the total~$\eta_{NLI}$ being added every~$t$ to the sum of provisioned LPs'~$\eta_{NLI}$.~The LPs are then added to~$L_{i,j,t+1}$, defined as a set of configured LPs between nodes~$(i,j)$ for the next planning period.

After additions are completed, we also check on the number of empty frequency slots in the link.~If more than 75\%~of the frequency slots have been filled up in any of the given links, it warns operators about upcoming frequency slot saturation in the configured fiber pair so that they plan a localized physical upgrade by exploring either additional bands or utilizing available dark fiber pairs\cite{Virgillito2019}.

\begin{table}[htbp!]
\scriptsize
\begin{tabular}{|c|c|c|c|c|c|} 
\hline
\textbf{Network} &
  \textbf{\begin{tabular}[c]{@{}c@{}}Nodes\\Links\\Demands\end{tabular}} &
  \textbf{\begin{tabular}[c]{@{}c@{}}Span\\variation\\(km)\end{tabular}} &
  \textbf{\begin{tabular}[c]{@{}c@{}}Noise\\Figure\\(dB) \supercite{Virgillito2019}\end{tabular}} &
  \textbf{\begin{tabular}[c]{@{}c@{}}Node\\Deg.(min/\\avg/max)\end{tabular}} \\ \hline
\textbf{Germany 17} &
  \textbf{17/26/272} &
  \textbf{30-120} &
  \textbf{4.3} &
  \textbf{2/3.05/6} \\ \hline
\textbf{US Abilene} &
  \textbf{12/15/132} &
  \textbf{20-100} &
  \textbf{4.3} &
  \textbf{1/2.5/4} \\ \hline
\end{tabular}
\caption{Reference topology information\supercite{phynwinfo}}
\label{tab:refnw}
\end{table}
Using the discussed flow, we setup a multi-period network study for a 17 node German backbone network with single mode fibers having heterogeneous span lengths\supercite{phynwinfo}.~We assume single fiber pair C-Band operation with variable gain EDFAs having constant noise figure as shown in Tab.~\ref{tab:refnw}.~The topological details, initial offered traffic and results for other topologies are also made publicly available\supercite{phynwinfo}.
\section{Results and Discussion}
We evaluate \textit{Scheme~1} and \textit{Scheme~2} in aforementioned traffic scenarios.~It must be noted that an additional scheme, where all BVTs are placed according to their maximum possible datarate has been evaluated in previous work\cite{Patri20}, compared to which, both \textit{Scheme~1} and \textit{Scheme~2} provide a higher network throughput.~For expected traffic growth shown in Fig.~\ref{fig:expgrowth}, we see that offered traffic can be satisfied by both schemes till 2025.~After which, we observe that the overall expected throughput for \textit{Scheme~2}~(dashed line with squares) cannot meet the offered traffic.~The cause may be attributed to the lack of additional frequency slots on some links, which cause additional LPs~($n_{c}$) found in Eq.~\ref{eq:ilp} to remain unprovisioned.~This translates to a loss of approximately 40 Tbps expected throughput at 2030.~In the same scenario, we see that \textit{Scheme~1} allows us to meet offered traffic without the need of a network upgrade for the entire~$t$.
\begin{figure}[htbp!]
\centering
 \includegraphics[width=0.45\textwidth,height=11em]{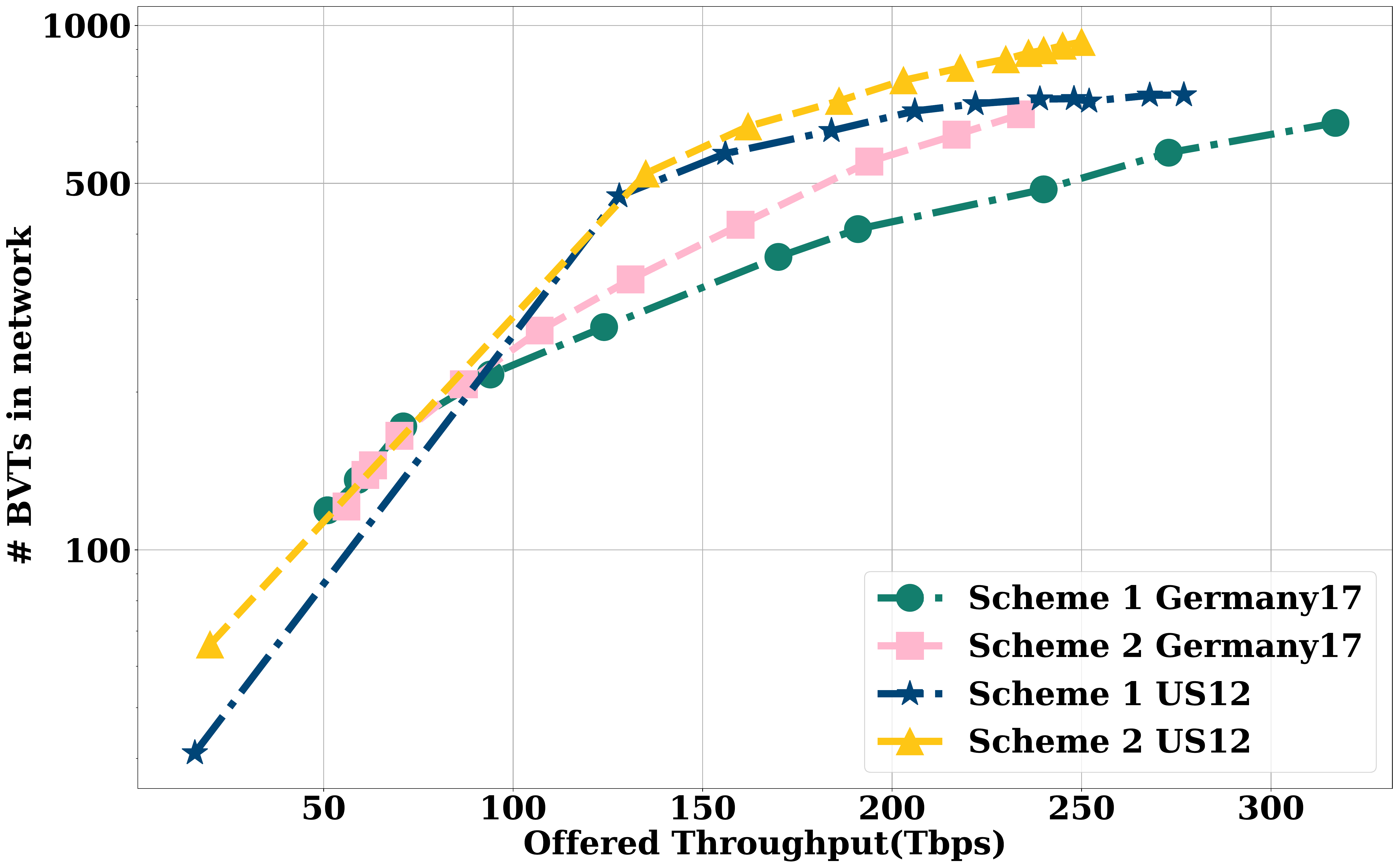}
\caption{BVTs vs Throughput for Germany 17 and US 12}
\label{fig:lpvstraff}
\end{figure}
\vspace{0.5em}
Similarly, for unexpected traffic growth~(shown in in Fig.~\ref{fig:unexpgrowth}), upgrading provisioned LPs using \textit{Scheme~1} enables the operators to utilize the C-Band for five additional years.~However, \textit{Scheme~1} also shows a downward trend in provisioned LPs, beginning from 2029.~We infer that in order to meet the offered traffic, network operators would have to upgrade their network, using new equipment and fibers or by exploring additional bands.~The question then arises as to when must the operator upgrade their network.~In the evaluated topology, we observe that three links cross the occupied slot threshold of 75\%~in the year 2027, which is when additional capacity must be planned for.~Post upgrade, the offered traffic can be met by the algorithm.

In Fig.~\ref{fig:lpvstraff} we see that \textit{Scheme~1} achieves a 40\%~increase in throughput for the same number of BVTs in Germany 17 topology.~Conversely, similar throughput can be achieved by \textit{Scheme~1} utilizing 25\%~lesser BVTs.~For the 12 node US Abilene topology~(US12), a 30\%~increase in throughput by using 18\%~lesser BVTs is observed.~The reduction may be because of long distance demands in the US12 topology, due to which offered traffic can only be met using more BVTs at lower modulation formats.~However, \textit{Scheme~1} still provides savings on the number of BVTs.

\section{Conclusions}
With increased global travel restrictions, core network traffic is bound to grow at a rate higher than initial forecasts.~Pragmatically, operators may use BVT over-provisioning to meet such surges in traffic. However, our proposed solution, evaluated with two network planning schemes, shows that a 40\%~overall increase in aggregate offered traffic for a realistic German core network can be carried up to five years, by efficiently planning the datarate and bandwidth usage of configured BVTs.~For higher traffic growth, an estimate on when to plan a physical network upgrade is also provided.
\section{Acknowledgements}
This work is partially funded by Germany's Federal Ministry of Education and Research under project OptiCON~(grant IDs \textbf{\#16KIS0989K} and \textbf{\#16KIS0991}).~The authors would also like to thank Dr.~Jose-Juan Pedreno-Manresa for fruitful discussions.~


\printbibliography
\vspace{-4mm}
\end{document}